# Some implications of the Hartree product treatment of the quantum nuclei in the ab initio Nuclear-electronic orbital methodology

Masumeh Gharabaghi[1] and Shant Shahbazian[2,*]

[1]Faculty of Chemical and Petroleum Sciences, Shahid Beheshti University, G. C. , Evin, Tehran, Iran, 19839, P.O. Box 19395-4716.

[2] Department of Physics, Shahid Beheshti University, G. C. , Evin, Tehran, Iran, 19839, P.O. Box 19395-4716.

E-mail:

(Shant Shahbazian) chemist_shant@yahoo.com

[*] Corresponding author

## Abstract

In this letter the conceptual and computational implications of the Hartree product type nuclear wavefunction introduced recently within the context of the ab initio non-Born-Oppenheimer Nuclear-electronic orbital (NEO) methodology are considered. It is demonstrated that this wavefunction may imply a pseudo-adiabatic separation of the nuclei and electrons and each nucleus is conceived as a quantum oscillator while a non-Coulombic effective Hamiltonian is deduced for electrons. Using the variational principle this Hamiltonian is employed to derive a modified set of single-component Hartree-Fock equations which are equivalent to the multi-component version derived previously within the context of the NEO and, easy to be implemented computationally.

## Keywords

Nuclear-electronic orbital methodology; non-Born-Oppenheimer; multi-component systems; Hartree product; muonic systems

# I. Introduction

In the last fifteen years several ab initio methodologies, with varying degree of success, have been developed aiming to solve Schrödinger's equation for atoms and molecules assuming both electrons and nuclei as quantum waves incorporating their kinetic energy operators simultaneously into the Hamiltonian [1-8]. These methodologies are beyond the usual ab initio procedures; within the Born-Oppenheimer (BO) paradigm electrons are treated as quantum waves and nuclei as clamped point charges, including exclusively the kinetic energy operators of electrons into the Hamiltonian [9]. The Nuclear-electronic orbital (NEO) is particularly a promising non-BO ab initio methodology since various types of the NEO wavefunctions have been proposed with varying degree of complexity [1,10-22]. Very recently, the NEO has also been successfully utilized for molecules containing exotic light quantum particles like the positively charged muons [23-26]. In this methodology certain types of nuclei or exotic light quantum particles are treated as quantum waves while adequate numbers of clamped nuclei are retained to avoid the complications emerging from treating total translational and rotational dynamics [27,28]. Similar to the case of the ab initio methodologies devised with the BO paradigm [9], the NEO methodology has also a "hierarchical" structure starting from the mean-field approximation, i.e. NEO Hartree-Fock (HF) [1]. The NEO-HF equations, in contrast to the orthodox HF equations, have a multi-component nature and a different set of the algebraic Roothaan-Hartree-Fock equations are constructed for each type of quantum particles that are solved simultaneously in the SCF procedure [1]. Accordingly, the spin-orbitals are attributed to both electrons and the nuclei treated as quantum waves, called hereafter quantum nuclei. In the next step in the hierarchical structure various types of correlations, e.g. electron-electron and electron-nucleus, are incorporated into the wavefunction designing various post-NEO-HF methods, e.g. the NEO multi-configurational HF [1], the NEO configuration interaction [1,21], the NEO many-body perturbation theory [22], and various variants of the explicitly correlated NEO-HF method [10-19].

The NEO-HF wavefunction is a product of Slater determinants for each type of quantum particles [1], however, when the overlap of nuclear spatial orbitals are negligible, which is often the case, the exchange and correlation effects attributed to the quantum nuclei are of no importance and the simpler Hartree product wavefunction may be used instead [20]. The total wavefunction in such case is the product of a Slater determinant of the electronic spin-orbitals and the Hartree product of nuclear spin-orbitals assuming that the quantum nuclei, irrespective of their type, are "distinguishable" quantum particles. This simplified wavefunction has been used in the NEO ab initio study of large molecular systems containing multiple quantum protons revealing a huge computational efficiency in comparison to the NEO-HF method [20]. In this letter the conceptual and computational implications of this simplified wavefunction is considered further.

## II. Reformulation of the NEO based on the Hartree product wavefunction: The modified single-component HF equations

Let's assume a molecular system containing $p$ types of quantum particles where type $1$ are electrons, there are $N_1$ electrons, and types $2$ to $p$ are distinguishable quantum nuclei while there are also $q$ number of clamped nuclei. The NEO Hamiltonian, satisfying time-independent Schrödinger's equation, $\hat{H}_{NEO}\Psi_{NEO} = E_{NEO}\Psi_{NEO}$, is as follows:

$$\hat{H}_{total} = \hat{H}_{NEO} + V_{classic}$$

$$\hat{H}_{NEO} = \left(\frac{-\hbar^2}{2m_1}\right)\sum_{i=1}^{N_1}\nabla_1^{i^2} + \sum_{n=2}^{P}\left(\frac{-\hbar^2}{2m_n}\right)\nabla_n^2 - \sum_{n=2}^{P}\sum_{i=1}^{N_1}\frac{Q_n}{|\vec{r}_n - \vec{r}_1^i|} + \sum_{n=2}^{P}\sum_{l\rangle n}^{P}\frac{Q_n Q_l}{|\vec{r}_n - \vec{r}_l|}$$

$$+ \sum_{i=1}^{N_1}\sum_{j\rangle i}^{N_1}\frac{1}{|\vec{r}_1^i - \vec{r}_1^j|} - \sum_{i=1}^{N_1}\sum_{\alpha=1}^{q}\frac{Z_\alpha}{|\vec{R}_\alpha - \vec{r}_1^i|} + \sum_{n=2}^{P}\sum_{\alpha=1}^{q}\frac{Q_n Z_\alpha}{|\vec{R}_\alpha - \vec{r}_n|}$$

$$V_{classic} = \sum_{\alpha=1}^{q}\sum_{\beta\rangle\alpha}^{q}\frac{Z_\alpha Z_\beta}{|\vec{R}_\alpha - \vec{R}_\beta|} \qquad (1)$$

In general various approximate methods have been introduced in order to solve the corresponding Schrödinger's equation [1,10-22]. One of possibilities is using the following trial normalized wavefunction assuming the Hartree product type wavefunction for the quantum nuclei (the spin variables have been neglected for brevity) within the context of the variational principle [20]:

$$\Psi_{trial} = \psi_1(\vec{r}_1^1,...,\vec{r}_1^{N_1})\prod_{n=2}^{P}\psi_n(\vec{r}_n) \qquad (2)$$

It is important to realize that at this stage the electronic part of the wavefunction, $\psi_1$, is not determined and in principle may be fully correlated, i.e. containing electron-electron correlation, however, the Hartree-type nuclear wavefunctions are composed of one-particle uncorrelated functions, $\psi_n$, thus no electron-nucleus or nucleus-nucleus correlations are contained in the total wavefunction from the outset [20]. Since the quantum nuclei are assumed to be "practically" distinguishable, no quantum "exchange" phenomenon is also expected for the quantum nuclei as well. In principle, the proposed NEO wavefunction may be used as a starting point to design variational based ab initio methods employing various approximate electronic wavefunctions; the simplest approximation is using a Slater determinant and then determining the electronic and nuclear one-particle wavefunctions from the variational principle. In their paper [20], pursuing this line of reasoning, Auer and Hammes-Schiffer derived coupled HF and Hartree-like equations for electrons and the quantum nuclei, respectively:

$$\hat{f}_1(\vec{r}_1^1)\psi_1^i(\vec{r}_1^1) = \varepsilon_1^i \psi_1^i(\vec{r}_1^1) \qquad i = 1, ..., N_1$$

$$\hat{f}_n(\vec{r}_n)\psi_n(\vec{r}_n) = \varepsilon_n \psi_n(\vec{r}_n) \qquad n = 2, ..., p \qquad (3)$$

In the first equation, $i$ enumerates the spectra of eigenfunction and eigenvalue of the electronic equations, while $\hat{f}$ stands for the Fock operators which are defined for electrons and the quantum nuclei as follows:

$$\hat{f}_1(\vec{r}_1^1) = \hat{h}_1(\vec{r}_1^1) + \sum_{j=1}^{N_1/2}\left[2\hat{J}_1^j(\vec{r}_1^1) - \hat{K}_1^j(\vec{r}_1^1)\right] - \sum_{n=2}^{p}\hat{J}_n(\vec{r}_1^1)$$

$$\hat{f}_n(\vec{r}_n) = \hat{h}_n(\vec{r}_n) + \sum_{l=2, l \neq n}^{p}\hat{J}_l(\vec{r}_n) - 2\sum_{i=1}^{N_1/2}\hat{J}_1^i(\vec{r}_n) \qquad (4)$$

In these equations $\hat{h}_1$ and $\hat{h}_n$ are the one-particle electronic and nuclear Hamiltonians, respectively:

$$\hat{h}_1(\vec{r}_1^1) = \left(-\hbar^2/2m_1\right)\nabla_1^2 - \sum_{\alpha=1}^{q}\frac{Z_\alpha}{\left|\vec{r}_1^1 - \vec{R}_\alpha\right|}$$

$$\hat{h}_n(\vec{r}_n) = \left(-\hbar^2/2m_n\right)\nabla_n^2 + \sum_{\alpha=1}^{q}\frac{Z_\alpha Q_n}{\left|\vec{r}_n - \vec{R}_\alpha\right|} \qquad (5)$$

Also, $\hat{J}$s are the coulomb operators:

$$\hat{J}_1^j(\vec{r}_1^1) = \int d\vec{r}_1^2 \, \psi_1^{*j}(\vec{r}_1^2)\left(\frac{1}{\left|\vec{r}_1^1 - \vec{r}_1^2\right|}\right)\psi_1^j(\vec{r}_1^2)$$

$$\hat{J}_n(\vec{r}_1^1) = \int d\vec{r}_n \, \psi_n^*(\vec{r}_n)\left(\frac{1}{\left|\vec{r}_1^1 - \vec{r}_n\right|}\right)\psi_n(\vec{r}_n)$$

$$\hat{J}_l(\vec{r}_n) = \int d\vec{r}_l \; \psi^*_l(\vec{r}_l) \left(\frac{1}{|\vec{r}_n - \vec{r}_l|}\right) \psi_l(\vec{r}_l)$$

$$\hat{J}_1^i(\vec{r}_n) = \int d\vec{r}_1^1 \; \psi^{*i}_1(\vec{r}_1^1) \left(\frac{1}{|\vec{r}_n - \vec{r}_1^1|}\right) \psi_1^i(\vec{r}_1^1) \tag{6}$$

While, $\hat{K}_1^j$ is the electronic exchange operator acting as follows:

$$\hat{K}_1^j(\vec{r}_1^1) \psi_1^i(\vec{r}_1^1) = \left[\int d\vec{r}_1^2 \; \psi^{*j}_1(\vec{r}_1^2) \left(\frac{1}{|\vec{r}_1^1 - \vec{r}_1^2|}\right) \psi_1^i(\vec{r}_1^2)\right] \psi_1^j(\vec{r}_1^1) \tag{7}$$

It is timely to emphasize that in this derivation the spin state of all the quantum nuclei are assumed to be in high-spin state, i.e., all $\alpha$ or all $\beta$, while the electrons are assumed to be closed-shell though the extension to the open-shell systems is straightforward. Also, while the electronic one-particle wavefunctions are assumed to be orthonormal, the nuclear one-particle wavefunctions are formally just normal; imposing the orthogonally condition on the Hartree product type wavefunctions during the derivation of the Hartree equations triggers certain mathematical complications that have been discussed elsewhere [29-35]. However, this is not a real problem here since in the case of "localized" nuclei the overlaps of the nuclear one-particle wavefunctions are practically null. In computational implementation of equations (3) both the electronic and the nuclear one-particle wavefunctions are expanded using the Gaussian basis functions and the resulting algebraic Roothaan-Hartree-Fock equations have been solved through the SCF procedure [20]. Since the nuclear one-particle wavefunctions are much more localized than their electronic congeners, the exponents of the nuclear Gaussian functions are much larger than those used for the electronic Gaussian functions [1,21]. Essentially, one may claim that the localized nuclear one-particle wavefunctions are describing the "vibrational" motion of the quantum nuclei or in other words, quantum nuclei are oscillators. From the computational viewpoint, equations (3) are clearly $p$-th coupled equations and they must be solved simultaneously to deduce the total energy-optimized wavefunction, however, let's consider them from an alternative viewpoint herein.

Because of the product nature of the wavefunction, it may be interpreted "pseudo-adiabatically", and accordingly each nuclear one-particle wavefunction may be conceived as the wavefunction of a "quantum oscillator". Consequently, each quantum nucleus is assumed to be in a hypothetical external potential and the following set of equations are proposed to describe these oscillators:

$$\hat{H}_n^{ocs}\psi_n(\vec{r}_n) = \varepsilon_n\psi_n(\vec{r}_n)$$

$$\hat{H}_n^{ocs} = \left(-\hbar^2/2m_n\right)\nabla_n^2 + \hat{V}_n^{ext.}(\vec{r}_n) \quad (8)$$

Obviously, the external potential energy function, $\hat{V}_n^{ext.}$, is not in *a priori* known and must be "constructed" using proper parameters to reproduce the one-particle nuclear wavefunctions, $\psi_n(\vec{r}_n)$, gained from the mentioned variational calculations; a linear combination of the Gaussian functions are usually used as the nuclear basis set [1,21,36], so in general, equations (8) describe anharmonic quantum oscillators.

At this stage of development let's assume that proper parameters are known for each oscillator and proceed; since $\hat{V}_n^{ext.}$ is assumed to be known, one may derive $\prod_{n=2}^{p}\psi_n(\vec{r}_n)$ and the only unknown will be the electronic wavefunction. To derive the proper electronic wavefunction the variational integral is used:

$$E = \int d\tau_1 \psi_1^*(\vec{r}_1^1...\vec{r}_1^{N_1})\left[\int d\tau_n \left(\prod_{n=2}^{p}\psi_n^*(\vec{r}_n)\right)\hat{H}_{NEO}\left(\prod_{n=2}^{p}\psi_n(\vec{r}_n)\right)\right]\psi_1(\vec{r}_1^1...\vec{r}_1^{N_1}) \quad (9)$$

For brevity, $d\tau_1$ and $d\tau_n$ are used for the product of the differential volumes of all electronic and all nuclear variables, respectively. Since the nuclear one-particle wavefunctions are assumed to be known, the bracket in the middle of the integral may be calculated explicitly employing the Hamiltonian given in equation (1) and after some mathematical manipulations equation (9) is transformed as follows:

$$E = \int d\tau_1 \psi_1^*(\vec{r}_1^{\,1}...\vec{r}_1^{\,N_1}) \, \hat{H}_{eff}^{elec} \, \psi_1(\vec{r}_1^{\,1}...\vec{r}_1^{\,N_1})$$

$$\hat{H}_{eff}^{elec} = \left(\frac{-\hbar^2}{2m_1}\right)\sum_{i=1}^{N_1}\nabla_1^{i\,2} + \sum_{i=1}^{N_1}\sum_{j>i}^{N_1}\frac{1}{\left|\vec{r}_1^{\,i}-\vec{r}_1^{\,j}\right|} - \sum_{i=1}^{N_1}\sum_{\alpha=1}^{q}\frac{Z_\alpha}{\left|\vec{R}_\alpha-\vec{r}_1^{\,i}\right|} + \sum_{n=2}^{p}\left(\frac{-\hbar^2}{2m_n}\right)\int d\vec{r}_n \, \psi_n^*(\vec{r}_n)\left(\nabla_n^2\right)\psi_n(\vec{r}_n)$$

$$-\sum_{n=2}^{p}\int d\vec{r}_n \, \psi_n^*(\vec{r}_n)\left(\sum_{i=1}^{N_1}\frac{Q_n}{\left|\vec{r}_n-\vec{r}_1^{\,i}\right|}\right)\psi_n(\vec{r}_n) + \sum_{n=2}^{p}\int d\vec{r}_n \, \psi_n^*(\vec{r}_n)\left[\sum_{\alpha=1}^{q}\frac{Z_\alpha Q_n}{\left|\vec{R}_\alpha-\vec{r}_n\right|}\right]\psi_n(\vec{r}_n)$$

$$+\sum_{n=2}^{p}\sum_{l>n}^{p}Q_n Q_l \int\int d\vec{r}_n \, d\vec{r}_l \, \psi_n^*(\vec{r}_n)\psi_l^*(\vec{r}_l)\left[\frac{1}{\left|\vec{r}_n-\vec{r}_l\right|}\right]\psi_n(\vec{r}_n)\psi_l(\vec{r}_l) \quad (10)$$

Equation (10) is particularly interesting since one may conceive that the quantum nuclei have been "dissolved" in the effective electronic Hamiltonian.

In order to construct an explicit model for the effective electronic Hamiltonian, the simplest conceivable nuclear basis set, i.e., a single s-type Gaussian basis function, $\psi_n(\vec{r}_n) = (2\alpha_n/\pi)^{\frac{3}{4}}\exp\left[-\alpha_n\left|\vec{r}_n-\vec{R}_{n,c}\right|^2\right]$, which has also been used in some previous ab initio NEO-HF calculations [23-26,36], is employed herein. In this function the two parameters, to be determined for each quantum nucleus, are the exponent, $\alpha_n$, and the center of the location of the function in space, $\vec{R}_{n,c}$. The s-type Gaussian function is also the ground state eigenfunction of the 3D isotropic harmonic oscillator (HO) and the external potential energy functions in equations (8) is deduced as follows:

$\hat{V}_n^{ext.} = \left(2\alpha_n^2 \hbar^2 / m_n\right)\left|\vec{r}_n - \vec{R}_{n,c}\right|^2$. If one incorporates the s-type Gaussian function into equation (10), after some mathematical manipulations, the following effective HO-based Hamiltonian is derived (*erf* stands for the error function):

$$\hat{H}_{eff}^{HO} = \left(\frac{-\hbar^2}{2m_1}\right)\sum_{i=1}^{N_1}\nabla_1^{i\,2} + \sum_{i=1}^{N_1}\sum_{j>i}^{N_1}\frac{1}{\left|\vec{r}_1^i - \vec{r}_1^j\right|} - \sum_{i=1}^{N_1}\sum_{\alpha=1}^{q}\frac{Z_\alpha}{\left|\vec{R}_\alpha - \vec{r}_1^i\right|} + \sum_{n=2}^{p}\left(\frac{3\alpha_n \hbar^2}{2m_n}\right)$$

$$-\sum_{n=2}^{p}\sum_{i=1}^{N_1}\frac{Q_n}{\left|\vec{r}_1^i - \vec{R}_{n,c}\right|}\,erf\left[\left(2\alpha_n\right)^{\frac{1}{2}}\left|\vec{r}_1^i - \vec{R}_{n,c}\right|\right] + \sum_{n=2}^{p}\sum_{\alpha=1}^{q}\frac{Q_n Z_\alpha}{\left|\vec{R}_\alpha - \vec{R}_{n,c}\right|}\,erf\left[\left(2\alpha_n\right)^{\frac{1}{2}}\left|\vec{R}_\alpha - \vec{R}_{n,c}\right|\right]$$

$$+\sum_{n=2}^{p}\sum_{l>n}^{p}\frac{Q_n Q_l}{\left|\vec{R}_{n,c} - \vec{R}_{l,c}\right|}\,erf\left[\left(\frac{2\alpha_n \alpha_l}{\alpha_n + \alpha_l}\right)^{\frac{1}{2}}\left|\vec{R}_{n,c} - \vec{R}_{l,c}\right|\right] \qquad (11)$$

Manifestly, the first three terms are those of the familiar orthodox electronic Hamiltonian derived within the BO paradigm while the remaining four unprecedented terms originate directly from the quantum nuclei. The fourth term emerges from the kinetic energy operators of the quantum nuclei and is the sum of the mean kinetic energies of $p$-th 3D isotropic harmonic oscillators. Since for a harmonic oscillator: $\alpha = \left((mk)^{\frac{1}{2}}/2\hbar\right)$ ($k$ stands for the force constant) in the limit of the large mass the kinetic energy term vanishes: $\lim_{m_n \to \infty}\sum_{n=2}^{p}(3\hbar/4)\left(k_n/m_n\right)^{\frac{1}{2}} \to 0$. The next three remaining terms emerge from the interaction of the quantum nuclei with electrons, with clamped nuclei, and with each other, respectively. All these terms are the product of the orthodox Coulombic interaction and the error function thus the "effective interaction" of a quantum nucleus with other particles is "non-Coulombic"; interestingly, this potential energy function has also been used previously in the partitioning of the Coulombic operator into long and short ranges [37,38].

Manifestly, near $\vec{R}_{n,c}$, the error function "damps" the Coulombic term while far from $\vec{R}_{n,c}$, the product is practically indistinguishable from the Coulombic interaction. Based on the previous discussion on the HO, in the large mass limit the exponents of the Gaussian functions tend to infinity thus the error functions disappear from the effective interaction terms and all the terms reduce to the orthodox Coulombic interaction. This is also what expected intuitively since a larger exponent implies a more localized distribution of a quantum nucleus which resembles more a clamped nucleus and in the large mass limit this distribution is a Dirac delta function representing practically a clamped point charge. One may sum up and claim that in the large mass limit the effective HO-based Hamiltonian reduces to the known electronic Hamiltonian written within the context of the BO paradigm:

$$\lim_{m_n \to \infty} \hat{H}_{eff}^{HO} \to \hat{H}_{elec}^{BO} + V_{classic}$$

$$\hat{H}_{elec}^{BO} = \left(\frac{-\hbar^2}{2m_1}\right)\sum_{i=1}^{N_1}\nabla_1^{i2} - \sum_{i=1}^{N_1}\left(\sum_{\alpha=1}^{q}\frac{Z_\alpha}{\left|\vec{R}_\alpha - \vec{r}_1^i\right|} + \sum_{n=2}^{p}\frac{Q_n}{\left|\vec{r}_n - \vec{r}_1^i\right|}\right) + \sum_{i=1}^{N_1}\sum_{j>i}^{N_1}\frac{1}{\left|\vec{r}_1^i - \vec{r}_1^j\right|}$$

$$V_{classic} = \sum_{n=2}^{p}\sum_{l>n}^{p}\frac{Q_n Q_l}{\left|\vec{r}_n - \vec{r}_l\right|} + \sum_{n=2}^{p}\sum_{\alpha=1}^{q}\frac{Q_n Z_\alpha}{\left|\vec{R}_\alpha - \vec{r}_n\right|} \qquad (12)$$

As stressed, equation (10) is the starting point to design ab initio methods based on the variational principle, $\delta_{\psi_1,\{\alpha_n,\vec{R}_{n,c}\}}E = 0$. In the following, this is illustrated in the case of the effective BO-based Hamiltonian, $\hat{H}_{eff}^{HO}$, employing a Slater determinant for the electronic wavefunction. Accordingly, incorporating the Slater determinant into the variational integral and after some mathematical manipulations the following energy expression emerges for a closed-shell system:

$$E = 2\sum_{i=1}^{N_1/2} \int d\vec{r}_1^1 \psi_1^{i*}(\vec{r}_1^1) \hat{h}_1' \psi_1^i(\vec{r}_1^1)$$

$$+ \sum_{i=1}^{N_1/2} \sum_{j=1}^{N_1/2} \left[ 2\int d\vec{r}_1^1 \int d\vec{r}_1^2 \psi_1^{i*}(\vec{r}_1^1) \psi_1^{j*}(\vec{r}_1^2) \frac{1}{|\vec{r}_1^1 - \vec{r}_1^2|} \psi_1^i(\vec{r}_1^1) \psi_1^j(\vec{r}_1^2) \right.$$

$$\left. - \int d\vec{r}_1^1 \int d\vec{r}_1^2 \psi_1^{i*}(\vec{r}_1^1) \psi_1^{j*}(\vec{r}_1^2) \frac{1}{|\vec{r}_1^1 - \vec{r}_1^2|} \psi_1^j(\vec{r}_1^1) \psi_1^i(\vec{r}_1^2) \right] + U$$

$$\hat{h}_1' = \left[ \left( -\frac{\hbar^2}{2m_1} \right) \nabla_1^{1^2} - \sum_{\alpha=1}^{q} \frac{Z_\alpha}{|\vec{r}_1^1 - \vec{R}_\alpha|} - \sum_{n=2}^{p} \frac{Q_n}{|\vec{r}_1^1 - \vec{R}_{n,c}|} erf\left[ (2\alpha_n)^{1/2} |\vec{r}_1^1 - \vec{R}_{n,c}| \right] \right]$$

$$U = \sum_{n=2}^{p} \left( \frac{3\hbar^2}{2m_n} \right) \alpha_n + \sum_{n=2}^{p} \sum_{\alpha=1}^{q} \frac{Q_n Z_\alpha}{|\vec{R}_\alpha - \vec{R}_{n,c}|} erf\left[ (2\alpha_n)^{\frac{1}{2}} |\vec{R}_\alpha - \vec{R}_{n,c}| \right]$$

$$+ \sum_{n=2}^{p} \sum_{l>n}^{p} \frac{Q_n Q_l}{|\vec{R}_{n,c} - \vec{R}_{l,c}|} erf\left[ \left( \frac{2\alpha_n \alpha_l}{\alpha_n + \alpha_l} \right)^{\frac{1}{2}} |\vec{R}_{n,c} - \vec{R}_{l,c}| \right] \qquad (13)$$

The first three terms are similar to those derived from the usual electronic Hamiltonian within the BO paradigm apart from the fact that $\hat{h}_1'$, in contrast to $\hat{h}_1(\vec{r}_1^1)$ (see expression (5)) [39], contains a new type of one-electron non-Coulombic potential energy term emerging from the interaction of electrons and the quantum nuclei. All remaining terms gathered in $U$ are devoid of the electronic one-particle wavefunctions and any variation simply reduces to an optimization procedure similar to that used for the geometry optimization of clamped nuclei. The first analytical derivatives of energy with respect to $\alpha_n$ and $\vec{R}_{n,c}$ [40], or alternatively various numerical procedures maybe used for the optimization of these parameters. Thus, only variation with respect to the electronic one-particle functions is left to be considered, $\delta_{\{\psi_1^i\}} E = 0$; the mathematical details of the variation are similar to those used to derive the orthodox single-component HF equations and are not reiterated herein [39]. The resulting modified single-component HF equations are as follows:

$$\hat{f}_1'(\vec{r}_1^1)\psi_1^i(\vec{r}_1^1) = \varepsilon_1^i \psi_1^i(\vec{r}_1^1) \qquad i = 1,...,N_1$$

$$\hat{f}_1'(\vec{r}_1^1) = \hat{h}_1'(\vec{r}_1^1) + \sum_{j=1}^{N_1/2}\left[2\hat{J}_1^j(\vec{r}_1^1) - \hat{K}_1^j(\vec{r}_1^1)\right] \qquad (14)$$

These equations are comparable with the electronic part of the equations (3) and (4) while the coulomb and exchange operators are those defined in expressions (6) and (7). Evidently, *the solution of equations (14) plus the simultaneous optimization of the nuclear parameters, i.e., $\alpha_n$ and $\vec{R}_{n,c}$, are equivalent to the solution of the coupled equations (3) using a single fully optimized s-type Gaussian function as a nuclear basis set*. Equations (14) are solved for a fixed configuration of clamped nuclei and to deduce the equilibrium geometry, an extra geometry optimization of clamped nuclei must be done as usual. Equations (14) are quite similar to the usual HF equations and the available ab initio computational packages based on the BO paradigm maybe easily re-programmed to solve these equations. The only additional required ingredient is the evaluation of the one-electron integrals emerging from the non-Coulombic interaction of electrons and the quantum nuclei which have been derived analytically elsewhere [41]. The whole procedure is extendable further using a larger nuclear basis set, assuming anharmonic anisotropic oscillator instead of simple isotropic oscillator in equations (8), incorporating a linear combination of s-, p- and d-type Gaussian functions in equation (10) as will be considered in a future contribution.

## III. Conclusion and prospects

The NEO methodology has been originally designed to study intrinsically non-adiabatic processes, e.g. proton-coupled electron transfer, which are beyond the BO paradigm though this was proved not to be an easy task since it has been demonstrated that correct inclusion of the electron-nucleus correlation is a vital ingredient for such studies. An accurate and in the same time computationally efficient recovery of the electron-nucleus correlation is still to be devised though much progress in this direction has been made [10-22].

On the other hand, the NEO methodology has been also used to study molecular systems containing exotic particles like positrons [13,15,17,42], or the positively charged muons [23-26]. In the case of the muonic (and positronic) systems the use of the NEO methodology is inevitable from outset since no "safe" adiabatic background has been justified yet and thus any comprehensive "muon-specific computational chemistry" must be based on the NEO. In order to devise efficient ab initio procedures for the muonic molecules one needs to include both electron-electron as well as muon-electron correlation (muonic systems contain just a single muon thus the muon-muon exchange/correlation is absent totally). Inclusion of the electron-electron correlation into the present proposed scheme is straightforward and equation (10) is the basis to design any variational post-(modified-single-component-HF) ab initio procedure. Alternatively, as also considered within the context of the NEO [43-48], one may utilize the NEO density functional theory (NEO-DFT), and extend the present proposed scheme to the Kohn-Sham equations [20]. The lack of safe adiabatic background makes inclusion of the muon-electron correlation more subtle, therefore, the usual strategy to "tune" the explicitly correlated NEO-HF methods to reproduce the known adiabatic results on the nuclear distribution though applicable, is not in general trustable [10-19]. An alternative strategy is to adapt a more semi-empirical viewpoint and try to fix the exponents of the nuclear/muonic basis set on proper values instead of trying to deduce them variationally to avoid the well-known "over-localization" of the nuclear/muonic distribution [11,14,18,19]. In this viewpoint the reproduction of the available experimental data from the muonic molecules, e.g. hyperfine coupling constants, maybe utilized to derive "fine-tuned" NEO-DFT methods. All these directions are now under consideration in our lab and the results will be offered in future publications.

## Acknowledgments

The authors are grateful to Mohammad Goli, Michael Pak and Rohoullah Firouzi for detailed reading of a previous draft of this paper and their fruitful comments.

# References

[1]     S. P. Webb, T. Iordanov, S. Hammes-Schiffer, J. Chem. Phys. 117 (2002) 4106.

[2]     M. Cafiero, S. Bubin, L. Adamowicz, Phys. Chem. Chem. Phys. 5 (2003) 1491.

[3]     H. Nakai, Int. J. Quantum Chem. 107 (2007) 2849.

[4]     T. Kreibich, van Leeuwen R., E. K. U. Gross, Phys. Rev. A. 78 (2008)  022501 .

[5]     T. Ishimoto, M. Tachikawa, U. Nagashima, Int. J. Quantum Chem. 109 (2009) 2677.

[6]     S. Bubin, M. Pavanelo, W.-C. Tung, K. L. Sharkey, L. Adamowicz, Chem. Rev. 113 (2013) 36.

[7]     J. Mitroy, S. Bubin, W. Horiuchi, Y. Suzuki, L. Adamowicz, W. Cencek, K. Szalewicz, J. Komasa, D. Blume, K. Varga, Rev. Mod. Phys. 85 (2013) 693.

[8]     R. Flores-Moreno, E. Posada, F. Moncada, J. Romero, J. Charry, M. Díaz-Tinoco, S. A. González, N. F. Aguirre, A. Reyes, Int. J. Quantum Chem. 114 (2014) 50.

[9]     T. Helgaker, P. Jørgenson, J. Olsen, Molecular Electronic-Structure Theory, John Wiley & Sons, New York, 2000.

[10]    C. Swalina, M.V. Pak, A. Chakraborty, S. Hammes-Schiffer, J. Phys. Chem. A. 110 (2006) 9983.

[11]    A. Chakraborty, M. V. Pak, S. Hammes-Schiffer, J. Chem. Phys. 129 (2008) 014101.

[12]    A. Chakraborty, S. Hammes-Schiffer, J. Chem. Phys. 129 (2008) 204101.

[13]    M. V. Pak, A. Chakraborty, S. Hammes-Schiffer, J. Phys. Chem. A. 113 (2009) 4004.

[14]    C. Ko, M. V. Pak, C. Swalina, S. Hammes-Schiffer, J. Chem. Phys. 135 (2011) 054106.

[15]    C. Swalina, M. V. Pak, S. Hammes-Schiffer, J. Chem. Phys. 136 (2012) 164105.

[16]    A. Sirjoosingh, M. V. Pak, C. Swalina, S. Hammes-Schiffer, J. Chem. Phys. 139 (2013) 034102.

[17]    A. Sirjoosingh, M. V. Pak, C. Swalina, S. Hammes-Schiffer, J. Chem. Phys. 139 (2013) 034103.

[18]    A. Sirjoosingh, M. V. Pak, K. R. Brorsen, S. Hammes-Schiffer, J. Chem. Phys. 142 (2015)  214107.

[19]	K. R. Brorsen, A. Sirjoosingh, M. V. Pak, S. Hammes-Schiffer, J. Chem. Phys. 142 (2015) 214108.

[20]	B. Auer, S. Hammes-Schiffer, J. Chem. Phys. 132 (2010) 084110.

[21]	J. H. Skone, M. V. Pak, S. Hammes-Schiffer, J. Chem. Phys. 123 (2005) 134108.

[22]	C. Swalina, M. V. Pak, S. Hammes-Schiffer, Chem. Phys. Lett. 404 (2005) 394.

[23]	M. Goli, Sh. Shahbazian, Phys. Chem. Chem. Phys. 16 (2014) 6602.

[24]	M. Goli, Sh. Shahbazian, Phys. Chem. Chem. Phys. 17 (2015) 245.

[25]	M. Goli, Sh. Shahbazian, Phys. Chem. Chem. Phys. 17 (2015) 7023.

[26]	M. Goli, Sh. Shahbazian, Chem. Eur. J. 22 (2016) 2525.

[27]	B. T. Sutcliffe and R. G. Woolley, Phys. Chem. Chem. Phys. 7 (2005) 3664.

[28]	B. T. Sutcliffe and R. G. Woolley, Chem. Phys. Lett. 408 (2005) 445.

[29]	J. C. Slater, Phys. Rev. 32 (1928) 339.

[30]	J. A. Gaunt, Proc. Cambridge Phil. Soc. 24 (1928) 328.

[31]	S. M. Blinder, Am. J. Phys. 33 (1965) 431.

[32]	M. Levy, T.-S. Nee, R. G. Parr, J. Chem. Phys. 63 (1975) 316.

[33]	F. E. Harris, Int. J. Quantum Chem. 13 (1978) 189.

[34]	S. A. Perera, D. E. Bernholdt, R. J. Bartlett, Int. J. Quantum Chem. 49 (1994) 559.

[35]	J. Garza, J. A. Nichols, D. A. Dixon, J. Chem. Phys. 112 (2000) 1150.

[36]	M. Hashimoto, T. Ishimoto, M. Tachikawa, T. Udagawa, Int. J. Quantum Chem. 116 (2016) 961.

[37]	J. P. Dombroski, S. W. Taylor, P. M. W. Gill, J. Phys Chem. 100 (1996) 6272.

[38]	R. D. Adamson, J. P. Dombroski, P. M. W. Gill, Chem. Phys. Lett. 254 (1996) 329.

[39]	A. Szabo, N. S. Ostlund, Modern Quantum Chemistry: Introduction to Advanced Electronic Structure Theory, Dover Publications Inc., New York, 1996.

[40]	Y. Yamaguchi, Y. Osamura, J. D. Goddard, H. F. Schaefer, A New Dimension to Quantum Chemistry: Analytic Derivative Methods in Ab Initio Molecular Electronic Structure Theory, Oxford University Press, Oxford, 1994.

[41]	P. M. W. Gill, R. D. Adamson, Chem. Phys. Lett. 261 (1996) 105.

[42]	P. E. Adamson, X. F. Duan, L. W. Burggraf, M. V. Pak, C. Swalina, S. Hammes-Schiffer, J. Phys. Chem. A. 112 (2008) 1346.

[43]	M. V. Pak, A. Chakraborty, S. Hammes-Schiffer, J. Phys. Chem. A. 111 (2007) 4522.

[44]   A. Chakraborty, M. V. Pak, S. Hammes-Schiffer, Phys. Rev. Lett. 101 (2008) 153001.

[45]   A. Chakraborty, M. V. Pak, S. Hammes-Schiffer, J. Chem. Phys. 131 (2009) 124115.

[46]   A. Sirjoosingh, M. V. Pak, S. Hammes-Schiffer, J. Chem. Theory Comput. 7 (2011) 2689.

[47]   A. Sirjoosingh, M. V. Pak, S. Hammes-Schiffer, J. Chem. Phys. 136 (2012) 174114.

[48]   T. Culpitt, K. R. Brorsen, M. V. Pak, S. Hammes-Schiffer, J. Chem. Phys. 145 (2016) 044106.